\documentclass[prc,aps,showpacs,showkeys,nofootinbib]{revtex4}
\usepackage{epsfig}
\usepackage{graphicx}
\begin{document}
\date{\today}
\title{{\bf Exact and approximate 
many-body dynamics with stochastic one-body density matrix
evolution}}

\author{Denis Lacroix}
\address{Laboratoire de Physique Corpusculaire,\\
ENSICAEN and Universit\'e de Caen,IN2P3-CNRS,\\
Blvd du Mar\'{e}chal Juin,\\
14050 Caen, France}

\begin{abstract}
We show that the dynamics of interacting fermions can be exactly 
replaced by a quantum jump 
theory in the many-body density matrix space. 
In this theory, jumps occur between densities formed of
pairs of Slater determinants, $D_{ab}=\left| \Phi_a \right> \left< \Phi_b \right|$, where each state 
evolves according to the Stochastic Schr\"odinger Equation (SSE) given in ref. \cite{Jul02}.
A stochastic Liouville-von Neumann equation is derived as well as the associated Bogolyubov-Born-Green-Kirwood-Yvon
(BBGKY) hierarchy. Due to the specific form of the many-body density along the path, 
the presented theory is equivalent to a stochastic theory in  
one-body density matrix space, in which each
density matrix evolves according to its own mean field augmented by a
one-body noise. Guided by the exact reformulation, 
a stochastic mean field dynamics valid in the weak coupling approximation is proposed. 
This theory leads to an approximate treatment of 
two-body effects similar to the extended Time-Dependent Hartree-Fock (Extended TDHF) scheme.
In this stochastic mean field dynamics, statistical mixing can be 
directly considered and jumps occur on a coarse-grained 
time scale. Accordingly, numerical effort is expected to be significantly reduced for applications.
\end{abstract}


\pacs{ 24.10.Cn, 26.60.Ky, 21.60.Ka} 
\keywords{ stochastic theories, many-body dynamics}
\maketitle

\section{Introduction}

The purpose of this paper is to discuss the possibility to substitute the description  
of the
evolution of quantum interacting fermions by a
stochastic mean-field dynamics of one-body density matrices. In view of present 
computational capabilities, stochastic methods appear as a promising
tool to address exactly or approximately the problem of
correlated mesoscopic quantum systems such as nuclei, atomic clusters or
Bose-Einstein condensates. Mean field theories, i.e. Hartree-Fock theories, are
rarely able to describe the large variety of phenomena occurring in quantum systems. 
It is generally necessary to extend mean field theory by including the effect of
two-body correlations \cite{Goe82}. During the past decades, several
approximate stochastic theories have been proposed to describe strongly
interacting systems\cite{Kad62,Ayi80,Gra81,Bal81,Ayi88,Ayi99,Rei92,Rei92-2,Oni95}.
These approaches have in common that the noise is due to the
residual part of the interactions acting on top of the mean field. However
they generally differ on the strategy used to incorporate noise. In
some cases, the residual interaction is treated using statistical assumptions \cite
{Ayi80,Gra81}, while in other cases the interaction induces fluctuations in the
wave-packets either by random phase-shifts \cite{Bal81} or by quantum jumps
according to the Fermi-Golden rule \cite{Rei92,Oni95}. The influence of 
correlations is sometimes treated using the notion of stochastic
trajectories in the one-body density matrix space\cite{Ayi88,Ayi99}.
This latter is, among the different theories, the only 
one that has been applied to large amplitude
collective motions in the semi-classical limit \cite{Col04}. Recently,
its quantal version has been used to describe small amplitude collective vibrations in nuclei 
\cite{Lac01}. However the application of a stochastic approach to the
quantum many-body dynamics remains an open problem both from a numerical as well as  
a conceptual point of view \cite{Lac04,Lac98-2}.

In this work a different strategy is used to obtain a stochastic formulation
of the many-body problem. During the last ten years, many efforts have been
made using functional integral techniques \cite
{Lev80,Ker83,Neg88} to address the problem of nucleons in strong two-body
interactions. These theories provide an exact stochastic 
formulation of quantum
problems and lead to the so-called 
quantum Monte-Carlo methods \cite{Cep95}. Recent applications to nuclear physics have
shown that stochastic methods can be applied successfully to
describe the structure of nuclei \cite{Koo97}. These methods can also 
be applied to the description of dynamical properties \cite{Neg88}. 
However the self-consistent mean field does not generally play a
special role. Indeed the stochastic part is driven either by
the kinetic energy part of the Hamiltonian or by a fixed one-body potential in the case of
shell model Monte-Carlo calculations \cite{Koo97}.
Recently a new formulation \cite{Car01,Jul02} has been proposed that 
combines the advantages of the Monte-Carlo methods and of the mean field theories.
Application of functional integral theories are of great interest since they pave 
the way to a full implementation of the nuclear static and dynamical many-body problem 
using mean field theories in
a well defined theoretical framework. 
However the direct application of exact stochastic dynamics to realistic situations 
remains numerically impossible and proper approximations should be developed.

In the first part of the article, the functional integral method and the
associated Stochastic Schroedinger Equation (SSE) developed in \cite{Jul02} for
many-body and one-body wave functions are presented. The theory is formulated in the more
general framework of exact stochastic dynamics in the many-body and/or
in the one-body density matrix space. The link between the different
formulations is underlined. 
In a second part, guided by the exact stochastic theory, 
an extended mean field theory \cite{Lac04} taking into account two-body effects in the weak coupling 
regime is given in terms of a new stochastic one-body evolution.

\section{Introduction and discussion of stochastic methods}

Functional integral methods have been used for a long time to provide a
useful reformulation of complex quantum systems \cite{Lev80,Ker83} (for a
review see \cite{Neg88}). This method has been applied 
with success to describe static properties of nuclei \cite{Koo97}. 
However, it was seldom used for
dynamical problems. Recently, an alternative formulation of the path
integral representation has been obtained in which the mean field theory plays
a specific role. We consider a general
many-body system described by the wave function 
$\left| \Phi \right\rangle $ which evolves according 
to the Hamiltonian\footnote{Note that
three-body (or higher) interactions are not considered here.}: 
\begin{equation}
H=\sum_{ij}T_{ij}a_{i}^{+}a_{j}+\frac{1}{4}%
\sum_{ijkl}V_{ijkl}a_{i}^{+}a_{j}^{+}a_{l}a_{k} , 
\label{eq:hamil}
\end{equation}
where the first term corresponds to the kinetic part of the Hamiltonian
while the second part is the two-body
interaction. We use the
convention of \cite{Abe96} concerning the labelling of one and two-body
operators. We denote $V_{ijkl}=\left\langle ij\left| \widetilde{v}_{12}\right|
kl\right\rangle $ where $\widetilde{v}_{12}$ is the antisymmetrized two-body 
interaction.

\subsection{Action of a quadratic Hamiltonian on a Slater determinant}

In ref. \cite{Koo97}, the general strategy to obtain ground state properties
of a many-body system using Monte-Carlo methods is described. The new aspect 
developed in ref. \cite{Jul02} is the introduction a self-consistent mean field
before the application of functional integral. In that case, only the residual part of 
the interaction which is not taken into account in mean field Hamiltonian 
is treated stochastically. In this section, we summarize how 
a general two-body Hamiltonian applied to a Slater determinant can be 
separated into a mean field part and a residual two-body contribution. Details 
are given in ref. \cite{Jul02}. 

We consider a Slater determinant $\left| \Phi
\right\rangle $ defined as $\left| \Phi \right\rangle =\Pi _{\alpha
}a_{\alpha }^{+}\left| 0\right\rangle $ , where the single particle states $%
\left| \alpha \right\rangle $ may not be orthogonal. Starting from the Hamiltonian 
(\ref{eq:hamil}), we have 
\begin{equation}
H\left| \Phi \right\rangle =\left( H_{1}+H_{res}\right) \left| \Phi
\right\rangle ,
\label{eq:hhres}
\end{equation}
with 
\begin{equation}
H_{1}\left| \Phi \right\rangle =\left( E_{0}+\sum_{\alpha _{1}\overline{%
\alpha _{1}}}\left\langle \bar{\alpha}_{1}\left| h_{MF}\left( \rho
_{1}\right) \right| \alpha _{1}\right\rangle a_{\bar{\alpha}_{1}}^{+}a_{\hat{%
\alpha}_{1}}\right) \left| \Phi \right\rangle ,  \label{eq:h1}
\end{equation}
where we denote by $\left| \bar{\alpha}_{1}\right\rangle $ the particle
states (i.e. the unoccupied states) and where $\rho _{1}=\sum \left| \alpha
_{1}\right\rangle \left\langle \hat{\alpha}_{1}\right| $ is the one-body
density associated with $\left| \Phi \right\rangle $. The states $\left| \hat{%
\alpha}_{1}\right\rangle $ are defined by $\left\langle \hat{\alpha}%
_{1}\left| {}\right. \alpha _{2}\right\rangle =\delta _{\alpha _{1}\alpha
_{2}}$. In this expression, $h_{MF}\left( \rho _{1}\right) $ is the
mean field Hamiltonian  
\begin{equation}
\begin{array}{ll}
h_{MF}\left( \rho _{1}\right) & =T_{1}+\overline{v}\left( \rho _{1}\right)
\end{array}
.
\end{equation}
In this equation, $\overline{v}\left( \rho _{1}\right) =Tr_{2}\left( 
\widetilde{v}_{12}\rho _{2}\right) $ is the mean field potential where $%
Tr_{2}\left( .\right) $ denotes the partial trace on the second particle. In
equation (\ref{eq:h1}), we denote 
\begin{equation}
E_{0}=Tr\left( \rho _{1}h_{MF}\left( \rho _{1}\right) -\frac{1}{2}\rho _{1}%
\overline{v}\left( \rho _{1}\right) \right).
\end{equation}
In the single particle basis defined above, it could be shown that
\begin{equation}
\left\langle \bar{\alpha}_{1} \bar{\alpha}_{2} 
\left| \widetilde{v}_{12} \right| \alpha _{1} \alpha _{2} \right\rangle = -
\sum_{s,\alpha _{1}\alpha _{2}%
\bar{\alpha}_{1}\bar{\alpha}_{2}}\hbar \omega _{s}\left\langle \bar{\alpha}%
_{1}\left| O_{s}\right| \alpha _{1}\right\rangle \left\langle \bar{\alpha}%
_{2}\left| O_{s}\right| \alpha _{2}\right\rangle,
\label{eq:os}
\end{equation}
where $O_s$ is a one-body operator\footnote{ Following ref. \cite{Abe96}, 
we will sometimes make use of the identity $\widetilde{v}_{12}=-\sum_{s}\hbar 
\omega _{s}O_{s}^{1}O_{s}^{2}$  which is a compact notation for matrix elements 
and is 
only valid in the particle-hole basis. Here, we use the same notations as
in ref. \cite{Lac04} where "1" and "2" refers to the particles 
on which the operator is acting.}. 
Note that, the latter transformation of the two-body 
matrix elements is a particular case of the more general transformations given in ref.
\cite{Koo97}. When the single particle basis is not the particle-hole state of the 
Slater determinant, additional terms should be accounted for. 
Using this transformation, the residual part of the Hamiltonian is 
\begin{equation}
H_{res}\left| \Phi \right\rangle =\frac{1}{4}\sum_{s,\alpha _{1}\alpha _{2}%
\bar{\alpha}_{1}\bar{\alpha}_{2}}\hbar \omega _{s}\left\langle \bar{\alpha}%
_{1}\left| O_{s}\right| \alpha _{1}\right\rangle \left\langle \bar{\alpha}%
_{2}\left| O_{s}\right| \alpha _{2}\right\rangle a_{\bar{\alpha}_{1}}^{+}a_{%
\bar{\alpha}_{2}}^{+}a_{\hat{\alpha}_{1}}a_{\hat{\alpha}_{2}}\left| \Phi
\right\rangle.
\label{eq:resphi}
\end{equation}
In next section, this expression is the starting point to derive the stochastic
Schroedinger equations using functional integral techniques.

\subsection{Functional integrals and stochastic many-body dynamics}

Functional integrals methods applied to quantum fermionic systems in interaction 
\cite{Lev80,Ker83} lead to general stochastic formulations of
the quantum many-body problem. They however also lead to specific
difficulties. For instance, the semi-classical limit of the functional
integral does not give naturally the Hartree-Fock theory, but only to the
Hartree theory. The interesting idea proposed in \cite{Jul02} is to use
the functional integral already accounting for the fact that the
Hamiltonian is applied to a Slater determinant. In this case, only the
residual (2 particle-2 hole) part of the Hamiltonian is interpreted as a
source of noise. This procedure is summarized now.

We consider the evolution of the system during a small time-step $\Delta t$.
Denoting by $\left| \Delta \Phi \right\rangle $ the associated evolution, we have: 
\begin{equation}
\left| \Phi + \Delta \Phi \right\rangle =U\left( \Delta t\right) \left| \Phi
\right\rangle =\exp \left( \frac{\Delta t}{i\hbar }H\right) \left| \Phi
\right\rangle ,
\end{equation}
where $U\left( \Delta t\right) $ is the propagator associated to $H$. Using
the Hubbard-Stratonovitch \cite{Hub59,Str58} functional integral on the
residual part only, the exact propagator transforms into an integral
equation \cite{Jul02}:
\begin{equation}
\begin{array}{ll}
U\left( \Delta t\right) \left| \Phi \right\rangle = & \int{ d\overrightarrow{%
{\bf \sigma }}G\left( \overrightarrow{{\bf \sigma }}\right) \exp \left( 
\frac{\Delta t}{i\hbar }H_{1}+\Delta B\left( \overrightarrow{{\bf \sigma }}%
\right) \right) \left| \Phi \right\rangle} .
\end{array} 
\label{eq:hubstr}
\end{equation}
$H_{1}$ is given by equation (\ref{eq:h1}) while $\Delta
B \left( \overrightarrow{{\bf \sigma }}\right) $ is a one-body operator
written as 
\begin{equation}
\Delta B\left( \overrightarrow{{\bf \sigma }}\right)\left| \Phi \right\rangle  =\sum_{\alpha \bar{%
\alpha},s} \lambda_{s}\left\langle \overline{\alpha }\left| O_{s}\right| \alpha
\right\rangle \Delta W_{s}a_{\bar{\alpha}}^{+}a_{\hat{\alpha}}\left| \Phi
\right\rangle ,  \label{eq:db1}
\end{equation}
where 
\begin{equation}
\lambda_{s}=\sqrt{\omega _{s}}\left( 1+isgn\left( \omega _{s}\right) \right) /2
\label{eq:as}
\end{equation}
and $\Delta W_{s}=\sqrt{\Delta t}$ $\sigma _{s}$ with $\sigma _{s}$ the
component of the vector $\overrightarrow{{\bf \sigma }}$. In equation (\ref
{eq:hubstr}), $G\left( \overrightarrow{{\bf \sigma }}\right)= \Pi_s g( \sigma_s) $ represents
the product of normalized Gaussian probabilities of width $1$ for the $\sigma _{s}$ variables. 
As in other functional integral formulations, we recover that 
the original propagator associated to the exact evolution can be replaced 
by an ensemble of propagators that depend on $\overrightarrow{\sigma}$. 
Equivalently, in the limit of infinitesimal time step ($\Delta t\longrightarrow dt$), 
this equation can be interpreted as a stochastic Schroedinger 
equation for the initial state. Using the standard notation for stochastic 
processes in Hilbert space \cite{Gar85}, we have:  
\begin{equation}
\left| \Phi \right\rangle +\left| d\Phi \right\rangle =\exp \left( \frac{dt}{i\hbar }H_{1}+dB\left(
t\right) \right) \left| \Phi \right\rangle .  \label{eq:stonn}
\end{equation}
Here $\left| d\Phi \right\rangle $ has to be interpreted as a
stochastic wave function. Since eq. (\ref{eq:hubstr}) is exact, it shows that
the exact dynamics of a Slater determinant can be replaced by an average
over stochastic evolution operators. In this expression, $dB\left( t\right) $
is a stochastic operator which depends on the stochastic variable $dW_{s}$
according to equation (\ref{eq:db1})\footnote{%
Note that in the limit $\Delta t\longrightarrow dt$, 
$dW_{s}$ plays directly the role of the
Gaussian normalized stochastic variable, and the introduction of $\sigma
_{s} $ is not required.}. In order to obtain this equation, the
Ito rules for stochastic calculus have been used \cite{Gar85} with
\begin{equation}
dW_{s_{1}}dW_{s_{2}}=\delta _{s_{1}s_{2}}dt .  \label{eq:ito}
\end{equation}
Using the latter properties in combination with the expression of $dB\left(
t\right) $, we obtain an equivalent of the fluctuation-dissipation theorem
that gives the link between the stochastic operator and the residual part of the Hamiltonian: 
\begin{equation}
\frac{1}{2}dB\left( t\right) dB\left( t\right) \left| \Phi \right\rangle =+%
\frac{dt}{i\hbar }H_{res}\left( t\right) \left| \Phi \right\rangle .
\label{eq:flucdis}
\end{equation}

Expression (\ref{eq:stonn}) is of particular interest. Indeed, according to the Thouless
theorem \cite{Rin80,Bla86}, the application of an operator of the form (\ref
{eq:stonn}) to a Slater determinant gives another Slater determinant. 
Therefore, the evolution of correlated systems of fermions can be replaced by
stochastic evolutions of an ensemble of Slater determinants. 
Since each evolution can be solved with numerical techniques used in mean field theories, 
SSE offer a chance to solve exactly the dynamics of strongly interacting fermionic 
systems.   
This property has already been noted in several pioneering works \cite{Lev80,Ker83,Neg88}.
A very similar conclusion has been reached  for the description of interacting bosons 
using Monte-Carlo wave function techniques \cite{Car01}.
In this case and more generally in 
the context of the stochastic description of open quantum systems, jumps between wave-packets 
are generally described using differential stochastic dynamics in Hilbert space 
\cite{Ple98,Gar00,Bre02}. Then, 
the evolution of $\left| d\Phi \right\rangle$ is directly considered. 

The equivalent differential equation associated to the jump process described here 
can be obtained by developing the exponential in eq. (\ref{eq:stonn}) in powers of $dt$.
Using Ito rules, we obtain :
\begin{equation}
\left| \Phi \right\rangle + \left| d\Phi \right\rangle = \left( 1 + \frac{dt}{i\hbar }H_1 
+ \frac{ 1 }{ 2 } dB\left( t\right) dB\left( t\right)  + dB\left( t\right)
\right) \left| \Phi \right\rangle . 
\end{equation}    
Using equations (\ref{eq:hhres}) and (\ref{eq:flucdis}),  we finally obtain
a stochastic Schroedinger equation for
the many-body wave function: 
\begin{equation}
\left| d\Phi \right\rangle =\left( \frac{dt}{i\hbar }H+dB\left( t\right)
\right) \left| \Phi \right\rangle . \label{eq:stocmany}
\end{equation}
In the following, this equation is referred to as the {\it many-body SSE}.
Equation (\ref{eq:stocmany}) is strictly equivalent to (\ref{eq:stonn}) and thus 
preserves the Slater determinant nature of the states along the stochastic trajectory.
This might appear surprising due to the appearance of the complete 
Hamiltonian $H$ in eq. (\ref{eq:stocmany}). This is a specific aspect of the
stochastic many-body theory using Ito stochastic calculus. 
Indeed, although $H$ (which contains the complete
two-body interaction) drives the initial state out from the Slater determinant
space, the stochastic part of the equation of evolution compensates this effect exactly.  
The exponential form (\ref{eq:stonn}) or the differential form (\ref{eq:stocmany}) describe 
the same stochastic process. However, differential equations 
are generally easier to manipulate \cite{Ple98,Gar00,Bre02}.  

\subsection{Equivalent quantum jump for single particle states}

Up to now, we have introduced notions associated with the stochastic mechanics of
many-body wave functions. 
This formulation is of great interest for applications since 
the stochastic evolution of the many-body wave function can be replaced by
the stochastic evolution of its single particle components. For completeness 
the equivalent differential equation of single particle wave function is given below.
It has been shown in ref. \cite{Jul02} that equation (\ref{eq:stonn}) leads 
to the single particle equation of motion  
\begin{equation}
\left| d\alpha \right\rangle =\frac{dt}{i\hbar }h\left( \rho _{1}\right)
\left| \alpha \right\rangle +\sum_{s}\lambda_{s}\left( 1-\rho _{1}\right)
O_{s}\left| \alpha \right\rangle dW_{s},  
\label{eq:stocone}
\end{equation}
where $h\left( \rho _{1}\right) $ is a one-body operator given by 
\begin{equation}
h\left( \rho _{1}\right) =h_{MF}\left( \rho _{1}\right) -\frac{1}{2}\rho _{1}%
\overline{v}\left( \rho _{1}\right) .  \label{eq:h}
\end{equation}
Equation (\ref{eq:stocone}) will be referred to as the {\it one-body SSE}.
We would like to stress again that eq. (\ref{eq:stocmany}) and the set of
single particle evolutions (eq. (\ref{eq:stocone})) are strictly equivalent. 

In this section, we have summarized the equivalence between 
quantum jump approaches in many-body and one-body spaces of wave-packets 
in order to describe interacting fermions. An equivalent formulation 
in terms of density matrices is highly desirable 
to compare the exact treatment with other stochastic methods.      


\section{Density matrix formulation}

In the previous section, we have considered the stochastic formulation of the
many-body problem using stochastic Schroedinger equation. In this
approach, all trajectories start from a Slater determinant and follow a
stochastic path in the Slater determinant space. 
Stochastic theories can also be applied if the system is initially correlated. In this
case, it is helpful to generalize the theory by introducing density
matrices. It has been shown in ref. \cite{Jul02} that the many-body
density matrix $D\left( t\right)$ associated with the system at all times can
be properly described by the average over an ensemble of pairs of 
non-orthogonal Slater determinants state vectors,
\begin{equation}
D\left( t\right) =\overline{\left| \Phi _{a}\right\rangle \left\langle \Phi
_{b}\right| }, \label{eq:dadb}
\end{equation}
each of them evolving according to eq. (\ref{eq:stocmany}). Here, 
the average over the initial ensemble has been introduced. In that case, the
notion of a quantum jump between the wave functions is replaced by a quantum jump in
the space of Slater determinants pairs. In the following, 
the properties of Slater-determinant dyadics are recalled and a stochastic BBGKY
hierarchy \cite{Kir46,Bog46,Bor46} is derived.

\subsection{Slater determinant dyadics: notations}

Let us consider a many-body density formed of two distinct Slater
determinants 
\begin{equation}
D_{ab}=\left| \Phi _{a}\right\rangle \left\langle \Phi _{b}\right| ,
\label{eq:dab}
\end{equation}
in which each Slater determinant is an antisymmetrized product of not
necessarily orthogonal single particle states 
\begin{eqnarray}
\left\{ 
\begin{array}{ll}
\left| \Phi _{a}\right\rangle & =\Pi _{\alpha }a_{\alpha }^{+}\left|
0\right\rangle \\ 
\left| \Phi _{b}\right\rangle & =\Pi _{\beta }a_{\beta }^{+}\left|
0\right\rangle
\end{array}
\right. . 
\end{eqnarray}
{ Note that $D_{ab}$ is neither hermitian nor normalized. However, 
for convenience we will still call it a density matrix.}
Starting from the many-body density matrix, one can obtain the generalized  
$k-$body density matrix (denoted by $\rho _{1\ldots k}$ ) by taking successive partial
traces. Using the same notation as in \cite{Abe96}, we have: 
\begin{eqnarray}
\rho _{1\ldots k}^{(ab)}=Tr_{_{k+1\ldots ,A}} \left( D_{ab}\right) , 
\end{eqnarray}
where $A$ is the size of the system. One can obtain the expression of
density matrices in terms of single particle states of the two
Slater determinants by introducing the overlap matrix elements between single particle
states, denoted  by $f$. The matrix elements
of $f$ are defined by $f_{\beta _{i}\alpha _{j}}=\left\langle \beta
_{i}|\alpha _{j}\right\rangle $. For instance, the one-body density matrix
is:
\begin{equation}
\rho _{1}^{(ab)}={\it \ }\det \left( f\right) \sum_{\alpha _{i}\beta
_{j}}\left| \alpha _{i}\right\rangle f_{\alpha _{i}\beta
_{j}}^{-1}\left\langle \beta _{j}\right| \equiv \det \left( f\right)
u_{1}^{\left( ab\right) } .
\end{equation}
More generally the $k-$body density matrix is the
antisymmetrized product of single particle densities \cite{Low55} 
\begin{equation}
\rho _{1,\ldots ,k}^{(ab)}=\det \left( f\right) {\cal A}\left(
u_{1}^{(ab)}\ldots u_{k}^{\left( ab\right) }\right) ,  \label{eq:rhok}
\end{equation}
where ${\cal A}\left( .\right) $ corresponds to the antisymmetrization
operator. Introducing the two-body 
correlation operator defined by 
\begin{equation}
C_{12}^{(ab)}=\rho _{12}^{(ab)}-{\cal A}\left( u_{1}^{(ab)}\rho
_{2}^{(ab)}\right) , 
\end{equation}
we have $C_{12}^{(ab)}=0$ for any state defined by equation (\ref{eq:dab}).

\subsection{Stochastic evolution of many-body density matrices}

The BBGKY hierarchy \cite{Kir46,Bog46,Bor46} has been widely used as a
starting point in order to obtain approximations \cite{Abe96} on the
evolution of complex systems. 
Therefore, an equivalent hierarchy associated to the exact stochastic mean field deduced 
from functional integrals  
is highly desirable to specify the possible links with other theories.  
In this section, starting from the stochastic
Schroedinger equation for the many-body wave function, we give the associated
stochastic formulation of the BBGKY hierarchy.
In the stochastic many-body dynamics, we consider the quantum jump
between two different density matrices $D_{ab}$ and $D_{ab}^{\prime }$. 
Starting from $D_{ab}$ given by eq. (\ref{eq:dab}), there are transitions towards
another density matrix given by $D_{ab}^{\prime }=\left| \Phi _{a}+d\Phi
_{a}\right\rangle \left\langle \Phi _{b}+d\Phi _{b}\right| $. The rules for
transitions are directly obtained from the rules for the jumps in the
wave functions space:
\begin{eqnarray}
\left\{ 
\begin{array}{rr}
\left| d\Phi _{a}\right\rangle & =\frac{dt}{i\hbar }H\left| \Phi
_{a}\right\rangle +dB_{a}\left| \Phi _{a}\right\rangle \\ 
\left\langle d\Phi _{b}\right| & =-\frac{dt}{i\hbar }\left\langle \Phi
_{b}\right| H+\left\langle \Phi _{b}\right| dB_{b}^{+} . 
\end{array}
\right. \label{dfadfb}
\end{eqnarray}
with
\begin{eqnarray}
\begin{array}{rr}
dB_{a} & =\sum_{s}\lambda_{s}\sum_{\hat{\alpha}\overline{\alpha }}a_{\overline{%
\alpha }}^{+}\left\langle \overline{\alpha }\left| O_{s}\right| \alpha
\right\rangle a_{\hat{\alpha}}dW_{s_{a}} \\ 
dB_{b}^{+} & =\sum_{s} \lambda_{s}^{\ast }\sum_{\hat{\beta}\overline{\beta }}a_{
\hat{\beta}}^{+}\left\langle \beta \left| O_{s}\right| \overline{\beta }%
\right\rangle a_{\overline{\beta }}dW_{s_{b}} .
\end{array}
\end{eqnarray}
The notation $dW_{s_{a}}$ and $dW_{s_{b}}$ are introduced in order
to emphasize that stochastic variables associated respectively to $\left| \Phi
_{a}\right\rangle $ and $\left| \Phi _{b}\right\rangle $ are statistically
independent, i.e. 
\begin{equation}
dW_{s_{a}}dW_{s_{b}}=0 .
\end{equation}
This complete eq. (\ref{eq:ito}) verified both by $dW_{s_{a}}$ and $%
dW_{s_{b}}$. With these rules, the evolution of the many-body density
matrix along the stochastic path is  given by
\begin{equation}
dD_{ab}=\frac{dt}{i\hbar }\left[ H,D_{ab}\right]
+dB_{a}D_{ab}+D_{ab}dB_{b}^{+} . \label{eq:dba}
\end{equation}
This equation is a stochastic version of the Liouville-von
Neumann equation for the density matrix. The evolution of the $k-$body density
matrix can be directly derived from expression (\ref{eq:dba}) and one obtains:
\begin{equation}
d\rho _{1\ldots k}^{(ab)}=\frac{dt}{i\hbar }Tr_{k+1,\ldots ,A}\left( \left[
H,D_{ab}\right] \right) +d{\cal W}_{ab}^{k}  . \label{eq:dabsto}
\end{equation}
The additional term corresponds to the stochastic part acting on the $k$-body density matrix
evolution:
\begin{eqnarray}
d{\cal W}_{ab}^{k}=Tr_{k+1,\ldots ,A}\left(
dB_{a}D_{ab}+D_{ab}dB_{b}^{+}\right)  .
\end{eqnarray}
The first part of eq. (\ref{eq:dabsto}) is nothing but the standard
expression of the $k^{th}$ equation of the BBGKY hierarchy whose explicit
form can be found in review articles \cite{Cas90,Rei94,Abe96}. The equation 
of motion for the $k-$body density matrix in the framework of the 
stochastic
many-body theory proposed in ref. \cite{Jul02} corresponds to the standard BBGKY term augmented by a one-body
stochastic noise.

\subsection{Evolution of the one-body density matrix}

Starting from (\ref{eq:dabsto}), an explicit form of the 
one-body density evolution can be found. Since for any $D_{ab}$,
we have $C^{(ab)}_{12}=0$, the first term in eq. (\ref{eq:dabsto}) reduces
to:
\begin{equation}
\begin{array}{ll}
Tr_{2\ldots ,A}\left( \left[ H,D_{ab}\right] \right) & =Tr_{2}\left( \left[
H,\rho _{12}^{(ab)}\right] \right) \\ 
& =\left[ h_{MF}\left( u_{1}^{(ab)}\right) ,\rho _{1}^{(ab)}\right] .
\end{array}
\end{equation}
The stochastic part reads: 
\begin{eqnarray}
\begin{array}{ll}
Tr_{2,\ldots ,A}\left( dB_{a}D_{ab}\right) = & \sum_{s} \lambda_{s}\sum_{\alpha \bar{%
\alpha}}\left\langle \bar{\alpha}\left| O_{s}\right| \alpha \right\rangle \\ 
& \times Tr_{2,\ldots ,A}\left( a_{\bar{\alpha}}^{+}a_{\hat{\alpha}%
}D_{ab}\right) dW_{s_{a}} .
\end{array}
\end{eqnarray}
Let us introduce a complete single particle basis. For any state $\left|
i\right\rangle $ and $\left| j\right\rangle $ of the basis, we have : 
\begin{equation}
\begin{array}{ll}
\left\langle i\left| Tr_{2,\ldots ,A}\left( a_{\bar{\alpha}}^{+}a_{\hat{%
\alpha}}D_{ab}\right) \right| j\right\rangle & =Tr\left( a_{j}^{+}a_{i}a_{%
\bar{\alpha}}^{+}a_{\hat{\alpha}}D_{ab}\right) \\ 
& =\sum_{kl}\left\langle k\left| {}\right. \bar{\alpha}\right\rangle
\left\langle \hat{\alpha}\left| {}\right. l\right\rangle Tr\left(
a_{j}^{+}a_{i}a_{k}^{+}a_{l}D_{ab}\right) .
\end{array}
\end{equation}
Using the fermionic commutation rules on creation/annihilation operators 
together with the definition of the one-
and two-body density matrices, we obtain: 
\begin{equation}
\begin{array}{ll}
\left\langle i\left| Tr_{2,\ldots ,A}\left( a_{\bar{\alpha}}^{+}a_{\hat{%
\alpha}}D_{ab}\right) \right| j\right\rangle & =\sum_{kl}\left\langle
k\left| {}\right. \bar{\alpha}\right\rangle \left\langle \hat{\alpha}\left|
{}\right. l\right\rangle \left\langle li\left| \rho _{12}^{(ab)}\right|
kj\right\rangle \\ 
& +\sum_{l}\left\langle i\left| {}\right. \bar{\alpha}\right\rangle
\left\langle \hat{\alpha}\left| {}\right. l\right\rangle \left\langle
l\left| \rho _{1}^{\left( ab\right) }\right| j\right\rangle .
\end{array}
\end{equation}
Using the fact that $\rho _{12}={\cal A}%
\left( u_{1}\rho _{2}\right) $, we finally obtain: 
\begin{eqnarray}
\begin{array}{r}
Tr_{2,\ldots ,A}\left( dB_{a}D_{ab}\right) =\sum_{s} \lambda_{s}\left(
1-u_{1}^{(ab)}\right) O_{s}\rho _{1}^{(ab)}dW_{s_{a}} \\ 
+\sum_{s} \lambda_{s}Tr\left( u_{1}^{(ab)}\left( 1-\rho _{a}\right) O_{s}\right)
\rho _{1}^{(ab)}dW_{s_{a}} ,
\end{array}
\end{eqnarray}
where the one-body density $\rho _{a}$ associated with $%
\left| \Phi _{a}\right\rangle $ has been introduced. The same treatment can be performed for
the second part of the stochastic term and the evolution
of the one-body density matrix finally reads: 
\begin{equation}
\begin{tabular}{ll}
$d\rho _{1}^{(ab)}=$ & $\frac{dt}{i\hbar }\left[ h_{MF}\left(
u_{1}^{(ab)}\right) ,\rho _{1}^{(ab)}\right] +db_{1}^{(ab)}$ ,%
\end{tabular}
\label{eq:rho1sto}
\end{equation}
with 
\begin{equation}
\begin{array}{ll}
db_{1}^{(ab)}= & \sum_{s} \lambda_{s}\left( 1-u_{1}^{(ab)}\right) O_{s}\rho
_{1}^{(ab)}dW_{s_{a}}+\sum_{s} \lambda_{s}Tr\left( u_{1}^{(ab)}\left( 1-\rho
_{a}\right) O_{s}\right) \rho _{1}^{(ab)}dW_{s_{a}} \\ 
& +\sum_{s} \lambda_{s}^{\ast }\rho _{1}^{(ab)}O_{s}\left(
1-u_{1}^{(ab)}\right) dW_{s_{b}}+\sum_{s} \lambda_{s}^{\ast }Tr\left( O_{s}\left(
1-\rho _{b}\right) u_{1}^{(ab)}\right) \rho _{1}^{(ab)}dW_{s_{b}} .
\end{array}
\label{eq:dbsto}
\end{equation}
It is interesting to note that although the single particle states entering in $%
\rho _{1}^{(ab)}$ do not evolve according to mean field theory but according
to $h\left( \rho _{1}^{(ab)}\right)$ given by (\ref{eq:h}), the
deterministic part associated with the evolution of the one-body density
reduces to the standard mean field propagation. Eq. (\ref{eq:rho1sto})
points out the central role played by the mean field Hamiltonian in the
stochastic many-body theory. In particular, it shows that any evolution of
a correlated physical system submitted to a two-body interaction can be
replaced by a set of mean field evolutions augmented by a one-body noise. Finally, it
is worth noticing that expression \ (\ref{eq:rho1sto}) can 
alternatively be obtained by differentiating directly $\rho _{1}^{(ab)}=\det \left(
f\right) \sum_{\alpha _{i}\beta _{j}}\left| \alpha _{i}\right\rangle
f_{\alpha _{i}\beta _{j}}^{-1}\left\langle \beta _{j}\right| $. 

\subsection{k-body density evolution from one-body density}

The stochastic evolution transforms a pair 
of Slater determinants into another pair of Slater determinants. Thus, all the
information on a single stochastic trajectory is contained in the stochastic
evolution of the one-body density evolution in eq. (\ref{eq:rho1sto}).
Indeed, the evolution of the $k-$body density matrix can be directly
obtained from the relation (\ref
{eq:rhok}) which is valid all along the stochastic path. Using the Ito rules, we have 
\begin{equation}
\begin{array}{ll}
d\rho _{1 \ldots k}^{(ab)}= & d\left( \det \left( f\right) \right) {\cal A}%
\left( u_{1}^{\left( ab\right) }\ldots u_{k}^{\left( ab\right) }\right) \\ 
& +\det \left( f\right) \sum_{i}{\cal A}\left( u_{1}^{\left( ab\right)
}\ldots du_{i}^{\left( ab\right) }\ldots u_{k}^{\left( ab\right) }\right) \\ 
& +d\left( \det \left( f\right) \right) \sum_{i}{\cal A}\left( u_{1}^{\left(
ab\right) }\ldots du_{i}^{\left( ab\right) }\ldots u_{k}^{\left( ab\right)
}\right) \\ 
& +\det \left( f\right) {\cal A}\left( \sum_{i\neq j}u_{1}^{\left( ab\right)
}\ldots du_{i}^{\left( ab\right) }\ldots du_{j}^{\left( ab\right) }\ldots
u_{k}^{\left( ab\right) }\right) .
\end{array}
\label{eq:dddd}
\end{equation}
It can be checked that the terms which are linear in $dt$ correspond to the deterministic
part of eq. (\ref{eq:dabsto}). The latter expression is also useful
in order to have an explicit form of the stochastic noise to all order in $k$.
In expression (\ref{eq:dddd}), $d \left( \det \left(f\right) \right)$ 
is deduced from equations (\ref{dfadfb}). We have 
\begin{eqnarray}
d \det(f) &=& \left< d\Phi_b \left| \right. \Phi_a \right> +  
\left< \Phi_b \left| \right. d\Phi_a \right>  \\
&=& \left< \Phi_b \left| dB_a + dB^+_b \right| \Phi_a \right>,
\end{eqnarray}
which gives 
\begin{eqnarray}
d\det \left( f\right) =\sum_{s} \lambda_{s} Tr\left( u^{(ab)}
_{1}\left( 1-\rho _{a}\right) O_{s}\right) dW_{s_{a}}
+\sum_{s} \lambda_{s}^{\ast } Tr\left( O_{s}\left(
1-\rho _{b}\right) u^{(ab)}_{1}\right) dW_{s_{b}} . 
\label{eq:ddetf}
\end{eqnarray}
In addition, the equation on $du_{i}^{\left( ab\right) }$ is
deduced from (\ref{eq:rho1sto}). Altogether, we obtain 
\begin{equation}
\begin{array}{ll}
d{\cal W}_{ab}^{k}= & \sum_{i,s} \lambda_{s}\left[ \left( 1-u_{i}^{\left( ab\right)
}\right) O_{s}^{i}\right] dW_{s_{a }}\rho _{1\ldots k}^{\left(
ab\right) } \\ 
& +\rho _{1\ldots k}^{\left( ab\right) }\sum_{i,s} \lambda_{s}^{\ast }\left[
O_{s}^{i}\left( 1-u_{i}^{\left( ab\right) }\right) \right] dW_{s_{a}}
\\ 
& +\sum_{s} \lambda_{s}Tr\left( u_{1}^{\left( ab\right) }\left( 1-\rho _{a}\right)
O_{s}\right) dW_{s_{a }}\rho _{1\ldots k}^{\left( ab\right) } \\ 
& +\sum_{s} \lambda_{s}^{\ast }Tr\left( O_{s}\left( 1-\rho _{b}\right)
u_{1}^{\left( ab\right) }\right) dW_{s_{b}}\rho _{1\ldots k}^{\left(
ab\right) } .
\end{array}
\label{eq:dwk}
\end{equation}
Here, we introduced the notation $O^{i}_s$ to denote that the one-body operator $O_s$ is
applied to the particle $"i"$.  
The possibility to derive the evolution of $\rho _{1\ldots k}$ for all $k$ from the evolution of $\rho_1$
is an illustration of an attractive aspect of this theory. Indeed, since we are considering 
pairs of Slater determinants, all the information on the dynamics is contained in their 
one-body densities.  
This proves that the exact evolution
of the density matrix of a correlated system through a two-body Hamiltonian
can always be replaced by the average evolution of uncorrelated states
each of them evolving in the one-body space according to its own mean field
augmented by a one-body stochastic noise.

\subsection{Summary}

\begin{figure}[tbp]
\begin{center}
\epsfig{file=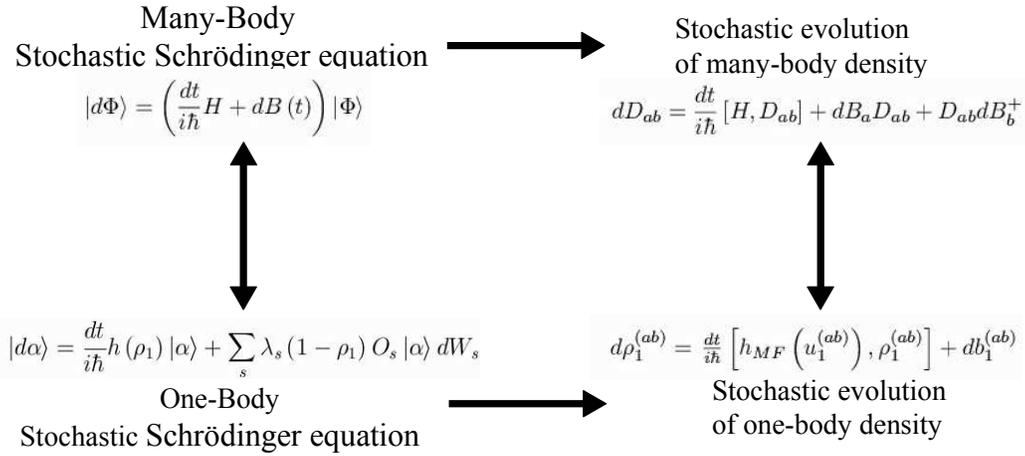,width=6.cm,angle=-90}
\end{center}
\caption{Summary of the four different ways of considering the exact
reformulation of the quantum many-body problem using stochastic mechanics.
The single arrow indicates that the density matrix formulation can be derived
from the stochastic Schroedinger equations. The double arrows show that 
for both wave function or density matrix formulations, there is a
strict equivalence between the many-body and the one-body stochastic equation of motion.}
\label{fig1sto}
\end{figure}

Functional integral methods are attractive since they provide a rather
transparent and systematic way of transforming the exact dynamics of a
correlated system into a stochastic mean field dynamics. In this work, we
have discussed the link between the different one-body and many-body SSE's on
one side and the stochastic one-body and many-body density evolution on the
other side. The equivalence and the relationship between the various ways of
considering stochastic mechanics are displayed in fig. \ref{fig1sto}. 

The exact stochastic formulation of the dynamics of complex systems provides a well
defined framework to introduce stochastic theories. However,
the stochastic dynamics as it is proposed is still rather cumbersome as far
as numerical applications are concerned. Indeed, due to the increasing number
of trajectories with the number of degrees of freedom, exact stochastic
many-body theories have only been applied to dynamics of rather schematic models\cite{Jul02}. 
With present computational facilities, there is no chance to
apply the exact theory to realistic mesoscopic systems and approximate
formulations are necessary. The stochastic theory provides however a natural
way to replace the dynamics of an interacting system by
one-body dynamical evolutions. In the following, we will transform 
the stochastic equation to account approximately for the correlation 
and reduce the numerical effort.   

\section{Approximate stochastic many-body dynamics}

A number of approximations of the many-body problem can be found in the literature. Among them,  
the mean field theory is certainly the most widely used. Correlations beyond the mean field are
often required to have a realistic description of dissipative aspects in
mesoscopic systems. A general strategy to obtain extensions of
the mean field dynamics consists in performing successive truncations of the
BBGKY hierarchy\cite{Cas90,Sur95,Lac04}. The first order truncation of the hierarchy
leads for instance to the standard mean field theory. An extension of
the mean field can be obtained by considering the first and second equations of
the hierarchy. This has led to different levels of approximations to the
nuclear many-body problem as for instance the so-called extended Time-Dependent
Hartree-Fock\cite{Won78,Orl79,Toh87} (for a recent review see ref. \cite{Lac04}). 
In the following, we will show that 
the stochastic evolution described previously can be adapted to a stochastic 
one-body theory for correlated systems equivalent to the
extended TDHF.

\subsection{Extended mean field dynamics}

Theories beyond mean field \cite{Wei80,Cas90} are valid when the dynamical effect of
the residual interaction is weak. In the weak coupling
regime, correlations can be treated perturbatively on top of the
mean field. These theories are valid under the assumption that different time
scales associated respectively to two-body collisions and to the mean field
propagation exist. Consider $\tau _{coll}$ the time scale for an in-medium two-body 
collision, and $\tau _{free}$ the time between two collisions.  In the weak
coupling approximation, one can assume that there exists a time interval $
\Delta t$ verifying the condition 
\begin{eqnarray}
\tau _{coll}\ll \Delta t \ll \tau _{free}.  \label{eq:time}
\end{eqnarray}
An estimate and a discussion of these time scales can be found in ref. 
\cite{Wei80,Kol95}. The physical picture to interpret the separation of
time scales is that each single particle state evolves according to the
average mean field and  rarely ''encounters'' a two-body collision. From the
many-body problem point of view, the role of the residual part of the interaction is 
to account for two-body collisions.

Besides time scales, extended TDHF remains a one-body theory. Indeed, it is assumed 
that part of the two-body 
correlations can be neglected and that the two-body density matrix can be instantaneously 
approximated by an antisymmetrized product of one-body density matrices 
($\rho _{12}(t)={\cal A} \left[ \rho_1(t) \rho _2(t) \right] $). 
This is of special interest
for practical applications since only one-body degrees of freedom are followed in time.
 
\subsection{Approximate stochastic dynamics}

In this section, we propose a formulation of extended one-body dynamics 
in terms of quantum jumps in the space of one-body density using the same hypothesis
as in extended TDHF. We start from
a system described at time $t_0$ by its one-body density given by 
\begin{equation}
\rho _{1}\left( t_0\right) =\sum_{\alpha }\left| \alpha \right\rangle
n_{\alpha }\left\langle \alpha \right| .
\label{rhostat}
\end{equation}
The system is assumed to be initially uncorrelated so that $\rho _{12} \left( t_0 \right)=
{\cal A}\left(\rho _{1}\left( t_0 \right)\rho _{2}\left( t_0 \right)\right)$. 
Let us now consider an ensemble of one-body density matrices, noted $\rho^{(n)}_1$ with initial
conditions $\rho^{(n)}_1(t_0)=\rho _{1}\left( t_0\right)$. The time interval $\Delta t$ is divided  
into $N$ time steps ($\Delta t = N \Delta s$) and at each time step, 
$\rho^{(n)}_1$ evolves according to its mean field augmented by a stochastic term
\begin{eqnarray}
 \Delta \rho^{(n)}_1= \frac{\Delta s}{i\hbar} \left[ h_{MF}\left( \rho^{(n)}_1 \right) , 
\rho^{(n)}_1 \right]+ \Delta K \left( \rho^{(n)}_1\right) .
\label{eq:rh1}
\end{eqnarray} 
However, contrary to the strategy of the previous section, and following the hypothesis of extended 
mean field theory, jumps are supposed to occur only 
once in the time interval $\Delta t$.  
For a jump occurring at a time $\tau=k \Delta s$, the 
stochastic term is written as 
\begin{eqnarray}
\begin{array} {lll}
\Delta K \left( \rho^{(n)}_1 \left( t \right)\right) = \delta_{t,\tau} &&
\left\{
\sum_{s} \lambda_{s}\left( 1-\rho^{(n)}_1 \left( t \right)\right) O_{s}\rho^{(n)}_1\left( t \right)
\Delta W_{s} \right.
\\
&&  \left.
+ \sum_{s'} \lambda_{s'}^{\ast }\rho^{(n)}_1\left( t \right) O_{s}
\left(1-\rho^{(n)}_1 \left( t \right)\right) \Delta W_{s'}
\right\} ,
\end{array} 
\end{eqnarray} 
where $\lambda_s$ and $O_s$ are defined in 
previous section, while $ \Delta W_s$ and $ \Delta W_{s'}$ are two independent 
Gaussian stochastic variables that follow Ito stochastic rules, with 
\begin{eqnarray}
\begin{array} {lll}
\overline{\Delta W_{s_1} \Delta W_{s_2}}  &=& \delta _{s_1s_2}\Delta s \\
\overline{\Delta W_{s'_1} \Delta W_{s'_2}}&=& \delta _{s'_1s'_2}\Delta s  .
\end{array}
\label{eq:itoap}
\end{eqnarray}
We consider the ensemble of trajectories with a quantum jump 
occurring at a specific time $\tau$.
$\overline{\rho^\tau_{12}}(t)$ denotes the  
two-body density obtained by averaging 
${\cal A} \left[\rho^{(n)}_1\rho^{(n')}_2\right]$
over these trajectories.  

Before the time $\tau$, all trajectories follow the same path corresponding to the mean field 
propagation with the initial condition $\rho _{1}\left( t_0\right)$. We note respectively 
$\rho^{mf}_1$ and
$U_{mf}$ 
the associated one-body density and propagator. We have 
\begin{eqnarray}
\rho^{mf}_1(t')=
U_{mf}\left( t^{\prime},t_0\right)\rho _{1}\left( t_0\right)U^+_{mf}\left( t^{\prime},t_0\right),
\end{eqnarray}
with 
\begin{eqnarray}
U_{mf}\left( t^{\prime},t_0\right)=T\exp \left( +\frac{1}{i\hbar }%
\int_{t_0}^{t^{\prime }}h_{mf}\left( \rho^{mf} _{1}\left( s\right) \right)
ds\right).
\end{eqnarray}
Using these definitions, the evolution between $\tau$ and $\tau + \Delta s$ of the product
$\left( \rho^{(n)}_1\rho^{(n')}_2 \right)$ is
\begin{eqnarray}
\Delta \left( \rho^{(n)}_1 \rho^{(n')}_2  \right) = \left( \Delta  \rho^{(n)}_1 \right)
\rho^{(n')}_2 +   \rho^{(n)}_1 \left( \Delta \rho^{(n')}_2 \right) +
\left( \Delta  \rho^{(n)}_1\right) \left( \Delta \rho^{(n')}_2 \right) .
\end{eqnarray}
Using expression (\ref{eq:rh1}) and Ito rules, we obtain
\begin{eqnarray}
\begin{array} {lll} 
\Delta \rho^{(n)}_1 \rho^{(n')}_2 (\tau) &=& \frac{\Delta s}{i \hbar} 
\left[ h_{mf}\left( \rho^{mf}_1(\tau ) \right)
+ h_{mf}\left( \rho^{mf}_2(\tau )\right), 
\rho^{mf}_1(\tau ) \rho^{mf}_2(\tau )  \right] \\
&& + \Delta K \left( \rho^{(n)}_1\right)\Delta K \left( \rho^{(n')}_2\right) 
+ \Delta K \left( \rho^{(n)}_1\right) + \Delta K \left( \rho^{(n')}_2\right).
\end{array}
\label{eq:inter1}
\end{eqnarray}
We have used the fact that, for all considered trajectories, 
no collision occurs before time $\tau$ leading to 
$\rho^{(n)}_1(\tau) =\rho^{(n')}_1(\tau) = \rho^{mf}_1(\tau )$. 
The last two terms  of equation (\ref{eq:inter1})  
do not contribute to the average evolution. We thus see that, in addition to the
mean field, an extra deterministic term will appear in the 
average evolution (\ref{eq:inter1}). Using equations (\ref{eq:itoap}),we have
\begin{eqnarray}
\begin{array} {lll}
\overline{
\Delta K \left( \rho^{(n)}_1\right)\Delta K \left( \rho^{(n')}_2\right) }&=&
\Delta s \sum_s  
(\lambda_{s})^2 \left( 1-\rho^{mf}_1 \left( \tau \right)\right)
\left( 1-\rho^{mf}_2 \left( \tau \right) \right)O^1_{s}
 O^2_{s}
\rho^{mf}_1\left( \tau \right) \rho^{mf}_2\left( \tau \right) \\
&+&\Delta s \sum_s  
(\lambda^*_{s})^2 \rho^{mf}_1\left( \tau \right) 
\rho^{mf}_2\left( \tau \right)O^1_{s}
O^2_{s}\left( 1-\rho^{mf}_1 \left( \tau \right)\right)
 \left( 1-\rho^{mf}_2 \left( \tau \right)\right).
\end{array}    
\end{eqnarray}
Using finally the fact that $\lambda^2_s = i\omega_s/2$, relation (\ref{eq:os}) 
and introducing the antisymmetrization operators, 
we obtain the average evolution
\begin{eqnarray}
 \overline{ \Delta {\cal A} \left( \rho^{(n)}_1   \rho^{(n')}_2 \right)(\tau ) }   = 
\frac{\Delta s}{i \hbar} 
\left[ h_{mf}\left( \rho^{mf}_1(\tau ) \right)
+ h_{mf}\left( \rho^{mf}_2(\tau )\right), {\cal A}\left( \rho^{mf}_1(\tau ) \rho^{mf}_2(\tau ) \right) \right]
+ \frac{ \Delta s }{ i \hbar } F_{12} \left( \tau  \right) . 
\end{eqnarray}
In this equation, $F_{12}$ reads
\begin{eqnarray}
\begin{array} {lll}
F_{12} (\tau ) &=& \left( 1-\rho^{mf} _{1}\left( \tau \right) \right) 
\left(1-\rho^{mf} _{2}\left( \tau \right) \right) \widetilde{v}_{12}
\rho^{mf} _{1}\left( \tau \right) \rho^{mf}
_{2}\left(  \tau \right) \\
&&- \rho^{mf} _{1}\left( \tau \right) \rho^{mf} _{2}\left( \tau \right)
\widetilde{v}_{12}
\left( 1-\rho^{mf} _{1}\left(  \tau  \right) \right) \left( 1-\rho^{mf} _{2}\left(
 \tau \right) \right) .
\end{array} 
\end{eqnarray}
As discussed in \cite{Lac98-2}, the effect of a 
single collision is expected to be weak 
during the time interval $\Delta t$ and we can assume that for all 
trajectories, the mean field 
propagation coincides with $U_{mf}$ after the jump. 
Therefore, the average density at the final time 
$t_f=t_0+\Delta t$ is given by:
\begin{eqnarray}
\overline{\rho^\tau_{12}}(t_f) = {\cal A} \left( \rho_1^{mf} (t_f) \rho_2^{mf} (t_f) \right)
+\frac{ \Delta s }{ i \hbar } U^{12}_{mf}\left( t_f,\tau \right) F_{12} \left( \tau  \right) 
U^{12+}_{mf}\left( t_f, \tau \right) ,
\end{eqnarray} 
where $U^{12}_{mf} = U^1_{mf} \otimes U^2_{mf}$. 
The complete average density $\overline{\rho_{12}}(t_f)$   
is obtained by summing different possible times $\tau$ for collisions:
\begin{eqnarray}
\overline{\rho_{12}}(t_f) = {\cal A} \left( \rho^{mf}_1 (t_f) \rho^{mf}_2 (t_f) \right)
+ \frac{ 1 }{ i \hbar }\int_{t_0}^{t_f}ds
U^{12}_{mf}\left( t_f, s \right) F_{12} \left( s  \right) 
U^{12+}_{mf}\left( t_f , s\right),
\end{eqnarray} 
where the limit $\Delta s \rightarrow ds$ has been taken.
This two-body density matrix corresponds to the standard mean field 
propagation augmented by the incoherent contribution of nucleon-nucleon
collisions entering generally in extended mean field theories \cite{Lac04}. 
 
As mentioned previously, an interesting aspect of extended TDHF is that it 
contains only one-body degrees of freedom. This can only be achieved 
by projecting correlation effects in the single particle space. 
In the stochastic dynamics presented here, this is equivalent to assume 
that the final two-body density can be approximated by ${\cal A}\left(\rho_1 (t_f) 
\rho_2 (t_f)\right)$    
where $\rho_1 (t_f)$ is given by 
\begin{eqnarray}
\rho_1 (t_f) = Tr_2\left( \overline{\rho_{12}}\left( t_f \right) \right).
\end{eqnarray}    
The density obtained in this way differs from the density propagated 
by the mean field alone and contains the effect of 
incoherent nucleon-nucleon collisions. The procedure can then be iterated 
using the new density as a starting point for future stochastic propagation.      

In this section, we have presented a method to include approximately two-body 
effects by means of a stochastic one-body theory. As in the exact formulation 
presented in the previous section, the stochastic theory can be equivalently formulated
as a stochastic Schroedinger equation. It is important to note that the numerical effort 
required for the approximate dynamics is expected to be much less than for the exact 
one at least for two reasons. The first one comes from the fact that quantum jumps 
occur on a "coarse-grained" time-scale. The second reason lies in the possibility of directly 
propagating densities formed by a statistical mixing (eq. (\ref{rhostat})) without invoking pairs 
of Slater determinants.
As a counterpart, we would like to mention that the approximate stochastic formulation 
has the same limitations at extended TDHF and can only be applied to problems for which 
the residual correlations are weak.    
  
\section{Conclusion}

The main result of our work is the proof that the exact dynamics of a
correlated system evolving through a two-body Hamiltonian can be replaced by
a set of stochastic evolutions of one-body density matrices where each density
evolves according to its own mean field augmented by a one-body noise. Guided 
by the exact stochastic formulation, an approximate stochastic mean field theory valid 
in the weak coupling limit  
is proposed. In this theory, jumps occur on a coarse-grained time scale. 

The alternative
stochastic formulation presented here does avoid some of the ambiguities
present in other stochastic theories. A first remarkable aspect
comes from the justification of the noise source. Indeed, since the starting
point of our work is an exact formulation of the many-body problem, the
noise has an unambiguous mathematical and physical interpretation.

In addition, from a practical point of view, it has clearly some
advantages. In all applications to quantum problems of extended mean field
theory, it has been shown that the memory effect is important (see
discussion in \cite{Lac98-2,Lac04}). This memory effect corresponds to the
non-local action in time of the past history collisions on the future
dynamics. Although the noise is Markovian, it accounts also for this
non-Markovian effect through the instantaneous average over trajectories. In
addition, as noted in ref. \cite{Lac98-2}, in order to apply stochastic
theories proposed in ref. \cite{Rei94,Ayi99} to large amplitude motions, one
should be able to guess what will be the important states in the future
evolution. This is in particular necessary to reduce the number of
trajectories. For instance, it has been guessed in ref. \cite{Rei92-2} that
jumps can be optimized due to the 2 particle-2 hole (2p-2h) nature of the
residual interaction. In the theory developed here, the system is driven
naturally towards the important states. Indeed, as can be
seen from eq. (\ref{eq:stocone}), these states are self-consistently
defined without ambiguity, and the 2p-2h character of
the residual interaction directly shows up in the stochastic part of the
propagator.

The exact treatment of the many-body problem 
with stochastic theories  is still
not possible for realistic large amplitude dynamics due to the required numerical effort. 
However, an alternative formulation of the stochastic theory
has been proposed in the second part of this article 
which should make the numerical applications easier. This stochastic 
theory provides a suitable framework for the description of interacting 
systems in the weak coupling regime. In particular, it keeps the advantages
discussed above and it is expected to significantly reduce the numerical 
efforts for practical applications. Such a theory could a priori be applied 
to nuclear systems where quantum and dissipative effects are important such as 
for instance giant resonances, fusion reactions or the thermalization 
in nuclear reactions.    

Finally we would  like to mention that an additional difficulty may be 
encountered due to the possible progressive entanglement of the initial state.
Indeed, starting from an initial simple state, the states propagated with 
stochastic Schroedinger equation will progressively become more 
complicated and fragmented over phase space. If such an entanglement occurs, 
the method proposed here might be very difficult or even impossible 
to use.
        
{\bf Acknowledgments}
The author thanks O. Juillet for helpful discussions during this work and 
S. Ayik, D. Durand and P. Van Isacker for a careful reading of the manuscript.

\end{document}